\documentclass[12pt]{article}

\begin{document}
\newcommand{\newc}{\newcommand}
\newc{\beq}{\begin{equation}}
\newc{\eeq}{\end{equation}}
\newc{\barr}{\begin{eqnarray}}
\newc{\earr}{\end{eqnarray}}
\newc{\ra}{\rightarrow}
\newc{\dfrac}{\displaystyle\frac}
\newc{\dint}{\displaystyle \int}
\newc{\half}{\frac{1}{2}}
\newc{\eq}[1]{(\ref{eq:#1})}
\newc{\eqs}[2]{(\ref{eq:#1},\ref{eq:#2})}
\newc{\etal}{{\it et al.}\ }
\newc{\etc}{{\it etc }\ }
\newc{\ie}{{\it i.e.}\ }
\newc{\eg}{{\it e.g.}\ }
\newc{\cA}{{\cal A}}
\newc{\cK}{{\cal K}}
\newc{\Ubar}{{\bar U}}
\newc{\Dbar}{{\bar D}}
\newc{\Ebar}{{\bar E}}
\newc{\nonum}{\nonumber}
\newc{\lab}[1]{\label{eq:#1}}
\newc{\vev}[1]{{<\!{#1}\!>\,}}
\newc{\gsim}{\stackrel{>}{\sim}}
\newc{\lsim}{\stackrel{<}{\sim}}
\newc{\kapi}{\frac{1}{\kap}}
\newc{\kz}[1]{(\kap z_{#1})}
\newc{\oc}{{\cal O}}
\newc{\vecl}{\vec{l}}
\newc{\lle}[3]{L_{#1}L_{#2}\Ebar_{#3}}
\newc{\lqd}[3]{L_{#1}Q_{#2}\Dbar_{#3}}
\newc{\udd}[3]{\Ubar_{#1}\Dbar_{#2}\Dbar_{#3}}

\topmargin-2.4cm
\begin{titlepage}
\begin{flushright}
ETH-TH/97-4
\end{flushright}
\vskip.3in
\begin{center}{\Large\bf
The Spectral Action Principle in Noncommutative\\
        Geometry and the Superstring }
\end{center}
\vskip.3in
\begin{center}{
 A. H. Chamseddine\\
 Theoretische Physik\\
 ETH-Z\"urich, CH-8093 } 
\end{center}
\vskip.5cm
\begin{center}
{\bf Abstract}
\end{center}
\begin{quote}
A supersymmetric theory in two-dimensions has enough data to 
define a noncommutative space thus making it possible to use
all tools of noncommutative geometry. In particular, we
apply this to the $N=1$ supersymmetric non-linear sigma model 
and derive an expression for the generalized loop space 
Dirac operator, in presence of
a general background, using canonical quantization. 
The spectral action principle is then used to determine
a spectral action valid for the fluctuations of the string modes.
\end{quote}
\end{titlepage}
\newpage
\medskip
It is now generally accepted that at very high energies, the structure of
space-time could not adequately be described by a manifold. Quantum
fluctuations makes it difficult to define localised points. The most
familiar example is string theory where points are replaced by strings, and
space-time becomes a loop space \cite{W82,W85, BR}. What has been lacking, up to now, are the
mathematical tools necessary to realize such spaces geometrically.
Fortunately, the recent advances in noncommutative geometry as formulated by
Connes \cite{ConnesBook} makes it possible to tackle such problems. The main advantage in
adopting Connes' formulation of noncommutative geometry is that the
geometrical data is determined by a spectral triple $({\mathcal{A}},\mathcal{%
H},D)$ where ${\mathcal{A}}$ is an algebra of operators, $\mathcal{H}$ a
Hilbert space amd $D$ a Dirac operator acting on $\mathcal{H}$. These ideas
have been successfully applied to simple generalizations of space-time such
as a product of discrete by continuos spaces . The results are very
encouraging in the sense that with a very simple imput one gets all the
details of the standard model including the Higgs mechanism, and the
unification of the Higgs fields with the gauge fields \cite{group}, as well as
unification with gravity \cite{ACAC, BKS} .

\medskip
Supersymmetric field theories in two-dimensions have enough data to define
noncommutative geometries \cite{CF, FG, FGR}. In  two-dimensions one can have $(p,q)$
supersymmetry as the left and right moving sectors could be split, giving
rise to various possibilities. The simplest possibilities are $N=1$ and $N={%
\frac{1}{2}}$ (i.e. $(1,1)$ and $(1,0)$ respectively). A good starting point
would be to consider various superconformal field theories and use the
noncommutative geometric tools to define geometric objects of interest. This
would be fruitful in cases such as orbifold compactifications where many
useful data is available to help define the noncommutative geometric space
completely. In this letter we shall adopt a slightly different framework
where the starting point is the supersymmetric non-linear sigma model in
two-dimensions \cite{Luis81}, with a general curved target space background. The conserved
supersymmetric charges satisfy the supersymmetry algebra \cite{W82}. Canonical
quantization would then change  these charges to Dirac operators over the
loop space $\Omega (M)$ where M is a Riemannian spin-manifold. The square of
a Dirac operator when restricted to reparametrization invariant
configurations gives the Hamiltonian of the system, as an elliptic
pseudo-differential operator. These operators could be used to write down a
spectral action as a function of the background fields, which gives the
low-energy effective action of string theory when the loops are shrunk to
points. At high-energies where oscillators are present the full spectral
action must be considered.

\medskip
The plan of this letter is as follows. First we give the essential
definitions needed to define a noncommutative space. Next we consider an $%
N=1 $ supersymmetric non-linear sigma model on a curved background with
torsion and construct the corresponding Dirac operator. The algebra is given
by the algebra of continous functions on the loop space, and the Hilbert
space is the Hilbert space $\mathcal{H}$ of states usually comprising two
sectors, the Ramond sector (R) with periodic boundary conditions for
fermions and Neveu-Schwarz (NS) with anti-periodic boundary conditions for
fermions. These data are then used to define a spectral action based on the
loop space Dirac operators, which will also give the Hamiltonian and
momentum operators in two-dimensions. We note that such considerations have
been performed before to determine the index of the loop space Dirac
operator (elliptic genus) but without torsion \cite{W87,JL}. Similar considerations were
also performed for the non-linear sigma models for point particles (with or
without torsion) \cite{Luis83, Davies, Braden},where the Hamiltonian of the system was 
determined. The
proposed action satisfies the constraints that it gives  at low-energies the string
effective action, and the partition function in the limit when the
background geometry is flat.

A starting point in defining noncommutative geometry \cite{ConnesBook} is the spectral 
triple $%
({\mathcal{A}},\mathcal{H},D)$ where ${\mathcal{A}}$ is a $*$ algebra of
bounded operators acting on a seperable Hilbert space $(\mathcal{H}$,
$D )$ is
a Dirac operator on $\mathcal{H}$ such that $[D,a]$ is a bounded operator
for arbitrary $a\in {\mathcal{A}}$. A K-cycle $\mathcal{H},D$ for $\mathcal{A%
}$ is said to be even if there exists a unitary involution $\Gamma $ on $%
\mathcal{H}$ such that $\Gamma a=a\Gamma $ for all $a\in {\mathcal{A}}$ and $%
\Gamma D=-D\Gamma $. Given a unital algebra $\mathcal{A}$ one can define the
universal differential algebra $\Omega ^{.}(\mathcal{A})=\oplus
_{n=0}^\infty \Omega ^n({\mathcal{A}})$ as follows: One sets $\Omega ^0({%
\mathcal{A}})={\mathcal{A}}$ and define $\Omega ^n({\mathcal{A}})$ to be the
linear space given by 
\begin{equation}
\Omega ^n({\mathcal{A}})=\left\{ \sum_ia_0^ida_1^i\cdots da_n^i\ :a_j^i\in {%
\mathcal{A}},\forall i,j\right\}
\end{equation}
where $d$ satisfies Liebnitz rule. We define a representation $\pi $ of $%
\Omega ^{.}({\mathcal{A}})$ on $\mathcal{H}$ by setting 
\begin{equation}
\pi \left( \sum_ia_0^ida_1^i\cdots da_n^i\right) =\sum_i\pi (a_0^i)[D,\pi ({%
a_1^i})]\cdots [D,\pi ({a_n^i})]
\end{equation}
Modulo the subtelty of moding the kernel of the representation $\pi $ out, it
is possible to define geometric objects such as distance, metric,
connection, curvature and so on \cite{CFFgravity}.

In physics, symmetry plays an important role, and it is important to
identify the symmetries associated with a noncommutative space. In gauge
theories the relevant symmetries are the internal gauge symmetries, and in
general relativity these are diffeomorphisms of the manifold. In noncommtative
geometry, it was shown
that these symmetries arise naturally as the automorphisms of the algebra ${%
\mathcal{A}}$, i.e. $Aut({\mathcal{A}})$ \cite{Connes}. Many  Dirac operators
associated with special geometries are known. 
Fluctuations induced under automorphisms of the algebra 
 would change the special   Dirac operators to the generic type.   
These are determined
by computing the one-form, $\pi (\rho ) =\sum_i a^i [D,b^i]$ so
that the new operators are of the form $D+ \pi (\rho )$.
The new Dirac operator would include information about 
 all the geometric invariants. It was conjectured in \cite{ACAC} that
the spectral action describes the dynamics of the fluctuations, and even
without knowing the exact form of the functional dependence of the
action on $D$, 
 a lot of information could
be extracted about the dynamics. Of course a well
 defined theory will
eventually have its spectral action in terms of a completely determined function.

In a different direction, in Witten's work on the relation between
supersymmetry and Morse theory \cite{W82}, it was shown that the supersymmetry charge
of a supersymmetric non-linear sigma model can be viewed as a Dirac operator
on an infinite dimensional loop space $\Omega (M)$ where $M$ is the target
space spin-manifold. In the case of $N=1$ supersymmetry there exists two
supersymmetry charges $Q_+$ and $Q_-$ satisfying 
\begin{equation}
\begin{array}{ccc}
Q_+^2 =H+P, &  &  \\ 
Q_-^2 =H-P, &  &  \\ 
\{ Q_+ ,Q_-\}=0 &  & 
\end{array}
\label{eq:2dsusy}
\end{equation}
If in the Hilbert space $\mathcal{H}$ a vacuum state is annihilated by $%
Q_{\pm}$ then it is also a vacuum state of the system. Restricting to states
satisfying $P=0$ would make the following identifications possible: 
\begin{equation}
\begin{array}{ccc}
Q_+= d+d^* , &  &  \\ 
Q_-= i(d-d^*), &  &  \\ 
H=dd^* +d^* d &  & 
\end{array}
\end{equation}
with $d^2 =d^{*2}=0$. The operators $d$ and $d^*$ are the differential
operator and its Hodge dual on $M$.

By considering different non-linear sigma models such as the $N= \frac{1}{2}$
model corresponding to heterotic strings, one obtains loop geometries which
are generalizations of $\Omega (M)$ allowing for gauge fields.

We start with the $N=1$ non-linear sigma model with background fields $%
G_{\mu \nu }\left[ \Phi \right] $ and $B_{\mu \nu }\left[ \Phi \right] $
which are symmetric and antisymmetric in $\mu \nu $ respectively. Here $\Phi
(\xi ,\theta _{+},\theta _{-})$ is a superfield with $\xi $ the coordinates
on the two-dimensional world sheet. The two-dimensional action is \cite{Luis81} 
\begin{equation}
I=\frac T2\dint d^2\xi d\theta _{+}d\theta _{-}\left( G_{\mu \nu }\left[
\Phi \right] +B_{\mu \nu }\left[ \Phi \right] \right) \left( D_{-}\Phi ^\mu
D_{+}\Phi ^\nu \right)  \label{action}
\end{equation}
whrere $T$ is the string tension and the component form of the superfield $%
\Phi ^\mu $ is 
\begin{equation}
\Phi ^\mu =X^\mu +i\theta _{+}\psi ^{\mu +}-i\theta _{-}\psi ^{\mu
-}+i\theta _{+}\theta _{-}F^\mu
\end{equation}
The operators $D_{\pm }$ are the supersymmetric derivatives (not to be
confused with Dirac operators which in this work will be denoted by $Q_{\pm
} $ ) and are given by 
\begin{equation}
\begin{array}{lrr}
D_{+}=\dfrac \partial {\partial \theta _{+}}-i\theta _{+}\partial _{+} &  & 
\\ 
D_{-}=\dfrac \partial {\partial \theta _{-}}-i\theta _{-}\partial _{-} &  & 
\\ 
\partial _{\pm }=\partial _0\pm \partial _1 &  & 
\end{array}
\end{equation}
We assume that the the two-dimensional reparametrization invariance has been
gauge fixed in the superconformal gauge, and the corresponding superghost
system has been added (we will come back to this point later). The
coordinates in two dimension are $\xi ^0=\tau $ and $\xi ^1=\sigma $ with $%
\tau \in R$ and $\sigma \in \left[ 0,2\pi \right] $. We will comment later on
the possibility of having more general backgrounds. The result of expanding
the action (\ref{action}) in component form, after eliminating the auxiliary
fields $F^\mu $ and performing the integration over the Grassmann
variables
 is well known,
and is given by \cite{Luis81} 
\begin{eqnarray}
I &=&\frac T2\dint d^2\xi \Bigl[ \left( G_{\mu \nu }[X]+B_{\mu \nu
}[X]\right) \partial _{-}X^\mu \partial _{+}X^\nu \Bigr.  \nonumber \\
&&\qquad +i\psi ^{a+}\left( \eta _{ab}\partial _{-}+\omega _{\mu
ab}^{+}\partial _{-}X^\mu \right) \psi ^{b+}+i\psi ^{a-}\left( \eta
_{ab}\partial _{+}+\omega _{\mu ab}^{-}\partial _{+}X^\mu \right) \psi ^{b-}
\nonumber \\
&&\qquad +\frac 12\psi ^{a+}\psi ^{b+}\psi ^{c-}\psi ^{d-}R_{cdab}^{+}[X] 
\nonumber \\
&&\qquad \left. +\frac i2\partial _{-}\left( B_{\mu \nu }[X]\psi ^{\mu
+}\psi ^{\nu +}\right) -\frac i2\partial _{+}\left( B_{\mu \nu }[X]\psi
^{\mu -}\psi ^{\nu -}\right) \right]  \label{eaction}
\end{eqnarray}
where 
\begin{eqnarray}
R_{\ \nu \rho \sigma }^{\mu \pm}  &=&\partial _\rho \Gamma _{\sigma \nu }^{\mu \pm
}+\Gamma _{\rho \kappa }^{\mu \pm }\Gamma _{\sigma \nu }^{\kappa \pm
}-\left( \rho \leftrightarrow \sigma \right) \\
\Gamma _{\nu \rho }^{\mu \pm } &=&\Gamma _{\nu \rho }^\mu \pm \frac 12H_{\
\nu \rho }^\mu \\
H_{\mu \nu \rho } &=&3\partial _{\left[ \mu \right. }B_{\left. \nu \rho
\right] }
\end{eqnarray}
The fermions with tangent indices $a,b,\ldots $ are related to those with
curved ones by: 
\begin{equation}
\psi ^a(\xi )=\psi ^\mu [X]e_\mu ^a[X]
\end{equation}
The varriation of the action (\ref{eaction}) gives the conserved currents $%
j_{\pm }$ \cite{Braden}
\begin{equation}
j_{\pm }=T\left( \psi ^{\mu \pm }G_{\mu \nu }\partial _{\pm }X^\nu \pm \frac
i6\psi ^{\mu \pm }\psi ^{\nu \pm }\psi ^{\rho \pm }H_{\mu \nu \rho }\right)
\end{equation}
Canonical quantization of the bosons give 
\begin{equation}
\left[ X^\mu (\sigma ,\tau ),P_\nu ({\sigma }^{\prime },\tau )\right]
=i\delta _\mu ^\nu \delta (\sigma -{\sigma }^{\prime })  \label{eq:xquant}
\end{equation}
while that for the fermions give (after replacing Poisson's brackets with
Dirac brackets) 
\begin{equation}
\left\{ \psi ^{a\pm }(\sigma ,\tau ),\psi ^{b\pm }({\sigma }^{\prime },\tau
)\right\} =\eta ^{ab}\delta (\sigma -{\sigma }^{\prime })
\end{equation}
The momentum $P_\mu (\sigma ,\tau )$ is related to the other fields through
the equation: 
\begin{equation}
P_\mu =T\left( G_{\mu \nu }\partial _0X^\nu +\frac i2\psi ^{a+}\psi
^{b+}\omega _{\mu ab}^{+}+\frac i2\psi ^{a-}\psi ^{b-}\omega _{\mu
ab}^{-}\right)
\end{equation}
where $\omega _{\mu ab}^{\pm }$ is related to the spin-connection $\omega
_{\mu ab}$ by 
\begin{equation}
\omega _{\mu ab}^{\pm }=\omega _{\mu ab}\pm \frac 12H_{\mu ab}
\end{equation}
and the spin-connection is defined through the condition 
\begin{equation}
\frac \delta {\delta X^\mu }e_\nu ^a[X]-\Gamma _{\mu \nu }^\rho
[X]e_\rho ^a[X]-\omega _{\mu \ b}^{\ a\ }[X]e_\nu ^b[X]=0
\end{equation}
The Dirac-Ramond operators $Q_{\pm }$ are defined as the integrals of the
currents $j_{\pm }$: 
\begin{equation}
Q_{\pm }=\int d\sigma j_{\pm }
\end{equation}
It proves more useful to rotate the fermions to the chiral basis \cite{Luis83}:  
\begin{equation}
\begin{array}{ccc}
\psi ^{a+} & =-\frac 1{\sqrt{2}}\left( \chi ^a+\overline{\chi }^a\right) & 
\\ 
\psi ^{a-} & =-\frac i{\sqrt{2}}\left( \chi ^a-\overline{\chi }^a\right) & 
\end{array}
\label{rotation}
\end{equation}
which will now satisfy the anticommutation relations: 
\begin{equation}
\left\{ \chi ^a(\sigma ,\tau ),\overline{\chi }^b({\sigma }^{\prime },\tau
)\right\} =\frac 12\eta ^{ab}\delta (\sigma -{\sigma }^{\prime })
\end{equation}
It is also useful to define the operators $Q$ and $\overline{Q}$ by 
\begin{equation}
\begin{array}{lcc}
Q_{+}=Q+\overline{Q} &  &  \\ 
Q_{-}=i\left( Q-\overline{Q}\right) &  & 
\end{array}
\label{qrotation}
\end{equation}
The final result is 
\begin{eqnarray}
Q &=&\int d\sigma \left( -\frac i{\sqrt{T}}\chi ^a(\sigma )e_a^\mu
[X]\left( \nabla _\mu +\frac{1}{3}H_{abc}\chi^b (\sigma )\chi^c (\sigma)
\right) \right.  \nonumber \\
&&\qquad +\left. \sqrt{T}\left( \overline{\chi }^a(\sigma )e_{\nu a}[X]-\chi
^a(\sigma )e_a^\mu [X]B_{\mu \nu }[X]\right) \frac{dX^\nu }{d\sigma }\right)
\label{Q} \\
\overline Q &=&\int d\sigma \left( -\frac i{\sqrt{T}}\overline{\chi }%
^a(\sigma )e_a^\mu [X]\left( \nabla _\mu
  +\frac{1}{3}H_{abc}\overline{\chi}^b(\sigma
  )\overline{\chi}^c(\sigma ) \right) 
\right.  \nonumber \\
&&\qquad +\left. \sqrt{T}\left( \chi ^a(\sigma )e_{\nu a}[X]-\overline{\chi }%
^a(\sigma )e_a^\mu [X]B_{\mu \nu }[X]\right) \frac{dX^\nu }{d\sigma }\right)
\label{Q'}
\end{eqnarray}
The covariant derivative $\nabla _\mu $ is defined by 
\begin{equation}
\nabla _\mu =\frac \delta {\delta X^\mu }+\omega _{\mu ab}[X]
\left( \chi ^a(\sigma )\overline{\chi }^b(\sigma )+\overline{\chi }%
^a(\sigma )\chi ^b(\sigma )\right)
\end{equation}
In determining the expressions for $Q$ and $\overline{Q}$ we have used the
quantization condition (\ref{eq:xquant}) to write 
\begin{equation}
P_\mu (\sigma )=-i\frac \delta {\delta X^\mu (\sigma )}  \label{moment}
\end{equation}

After a very lengthy calculation, the details of which will be given
somewhere else, one can show that 
\begin{equation}
Q^2=\overline{Q}^2=\frac 12P
\end{equation}
where the two-dimensional momentum $P$ is given by 
\begin{equation}
P=-i\dint d\sigma \frac{dX^\mu }{d\sigma }\nabla _\mu +2i\int d\sigma \chi
_a(\sigma )\frac{D\overline{\chi}^a}{D\sigma }
\end{equation}
and the covariant derivative $\frac D{D\sigma }$ is defined by 
\begin{equation}
\frac{D\overline{\chi }^a}{D\sigma }=\frac{d\overline{\chi }^a}{d\sigma }+%
\frac{dX^\mu }{d\sigma }\omega _{\mu \ b}^{\ a\ }[X]\overline{\chi }%
^b(\sigma )
\end{equation}
A lengthier calculation gives the anticommutator $\{Q,\overline{Q}\}=H$
where 
\begin{eqnarray}
H &=&-\frac 1{2T}\dint d\sigma \left[ \left( \nabla ^a\nabla _a+\omega _b^{\
ab}[X(\sigma )]\nabla _a+4\chi ^a(\sigma )\overline{\chi }^b(\sigma )\chi
^c(\sigma )\overline{\chi }^d(\sigma )R_{abcd}[X(\sigma )]\right) \right. 
\nonumber \\
&&\;\;\qquad \quad +\frac 23\left( \chi ^a(\sigma )\overline{\chi }^b(\sigma
)\overline{\chi }^c(\sigma )\overline{\chi }^d(\sigma )+\overline{\chi }%
^a(\sigma )\chi ^b(\sigma )\chi ^c(\sigma )\chi ^d(\sigma )\right) \nabla
_aH_{bcd}[X(\sigma )]  \nonumber \\
&&\qquad \qquad +\left( \chi ^b(\sigma )\chi ^c(\sigma )+\overline{\chi }%
^b(\sigma )\overline{\chi }^c(\sigma )\right) H_{\ bc}^a[X(\sigma )]\nabla _a
\nonumber \\
&&\qquad \qquad +\frac 13\left( \chi ^a(\sigma )\chi ^b(\sigma )\overline{%
\chi }^c(\sigma )\overline{\chi }^d(\sigma )+\chi ^a(\sigma )\overline{\chi }%
^b(\sigma )\overline{\chi }^c(\sigma )\chi ^d(\sigma )\right.  \nonumber \\
&&\qquad \qquad \qquad +\left. \overline{\chi }^a(\sigma )\overline{\chi }%
^b(\sigma )\chi ^c(\sigma )\chi ^d(\sigma )\right) H_{ab}^{\ \ e}[X(\sigma
)]H_{ecd}[X(\sigma )]  \nonumber \\
&&\qquad \qquad -2iT\left( \chi _a\frac{D{\chi }^a}{D\sigma }+\overline{\chi 
}_a\frac{D\overline{\chi }^a}{D\sigma }\right)  \nonumber \\
&&\qquad \qquad -2iT\left( \chi ^a\frac{D\chi ^b}{D\sigma }+\overline{\chi }%
^a\frac{D\overline{\chi }^b}{D\sigma }\right) B_{ab}[X(\sigma )]  \nonumber
\\
&&\qquad -2iTe_a^\mu [X(\sigma )]e_b^\nu [X(\sigma )]\left( \chi ^b(\sigma )%
\overline{\chi }^a(\sigma )+\overline{\chi }^b(\sigma )\chi ^a(\sigma
)\right) \left( \nabla _\rho B_{\mu \nu }-\nabla _\nu B_{\mu \rho }\right) 
\frac{dX^\rho }{d\sigma }  \nonumber \\
&&\qquad +2iTB_{\mu \nu }\frac{dX^\nu }{d\sigma }\nabla ^\mu -2i\left( \chi
^a(\sigma )\chi ^b(\sigma )+\overline{\chi }^a(\sigma )\overline{\chi }%
^b(\sigma )\right) H_{ab}^{\ \ c}[X(\sigma )]B_{c\nu }[X(\sigma )]\frac{%
dX^\nu }{d\sigma }  \nonumber \\
&&\left. \qquad \!\!\!\!\!-T^2\left( G_{\mu \nu }[X(\sigma )]+B_{\mu \kappa
}[X(\sigma )]B_{\ \nu }^\kappa [X(\sigma )]\right) \frac{dX^\mu }{d\sigma }%
\frac{dX^\nu }{d\sigma }\right]  \label{hamilton}
\end{eqnarray}
Throughout eq (\ref{hamilton}), curved indices are changed to tangent ones
with $e_a^\mu [X(\sigma )]$ and its inverse. It is clear that the
Hamiltonian is a second order elliptic operator in loop space, and that $P$
is the generator of reparametrizations on the circle. The supersymmetry
generators $Q_{\pm }$ are related to $Q$ and $\overline{Q}$ by (\ref
{qrotation}) . The target space coordinates $X^\mu (\sigma ,\tau )$ are
taken at time $\tau =0$. The dependence on time is governed by the equation: 
\begin{equation}
X^\mu (\sigma ,\tau )=e^{-\tau H}X^\mu (\sigma ,0)\,\,e^{\tau H}
\end{equation}
The advantage of adopting the canonical quantization is that one can use the
Fourier expansion of $X^\mu (\sigma )$. In the case of the closed superstring we
have (at $\tau =0$): 
\begin{equation}
X^\mu (\sigma )=X_0^\mu +\sum_{n>0}\frac 1{\sqrt{\pi nT}}\left( a_n^\mu \cos 
{n\sigma }+\tilde a_n^\mu \sin {n\sigma }\right)
\end{equation}
and the momentum $P_\mu $ of (\ref{moment}) is expressed as 
\begin{equation}
P_\mu =-i\left( \frac 1{2\pi }\frac \delta {\delta X_0^\mu }+\sum\limits_{n>0}%
\sqrt{\frac{nT}\pi }\left( \frac \delta {\delta a_n^\mu }\cos {n\sigma }%
+\frac \delta {\delta \tilde a_n^\mu }\sin {n\sigma }\right) \right)
\end{equation}
Similarly, the fermions $\chi ^a(\sigma )$ and $\overline{\chi }^a(\sigma )$
can be expanded in terms of oscillators: 
\begin{equation}
\begin{array}{ccc}
\chi ^a(\sigma )=\frac 1{\sqrt{2\pi }}\sum\limits_{r\in Z_0+\phi }\left(
c_r\cos {r\sigma }+d_r\sin {r\sigma }\right) &  &  \\ 
\overline{\chi }^a(\sigma )=\frac 1{\sqrt{2\pi }}\sum\limits_{r\in Z_0+\phi
}\left( \overline{c}_r\cos {r\sigma }+\overline{d}_r\sin {r\sigma }\right) & 
& 
\end{array}
\end{equation}
where $\phi =0$ for Ramond (R) boundary conditions (i.e. periodic) and $\phi
=\frac 12$ for Neveu-Schwarz (NS) boundary conditions. The quantization
conditions on the fermions imply that the only non-vanishing
anti-commutators are 
\begin{equation}
\begin{array}{ccc}
\{c_r^a,\overline{c}_s^b\}=2\delta _{rs}\eta ^{ab} & r, s \neq 0 &  \\  
\{ d_r^a, \overline{d}_s^b\}=2\delta_{rs}\eta^{ab} & 
& 
\end{array}
\end{equation}
The fermionic zero modes occur only in the R-sector, which satisfy the
anticommutation relations: 
\begin{equation}
\{c_0^a,\overline{c}_0^b\}=\eta ^{ab} 
\end{equation}
Therefore both $c_0^a+\overline{c}_0^a$ and $i(c_0^a-\overline{c}_0^a)$
generate Clifford algebras, and give rise to creation and annihilation operators for
the vacuum state.

This is not the full story. In obtaining the action, the
superparametrization invariance has been fixed, and the superghost
part must be
added to 
compensate for fixing the metric and gravitino.
This is well known and given by (see e.g. \cite{GSW} ): 
\begin{equation}
I^{(\mathrm{ghost)}}=-\frac 1{2\pi }\int d^2\xi d\theta _{+}d\theta
_{-}\left( BD_{-}C+\overline{B}D_{+}\overline{C}\right)
\end{equation}
where the fields $B$ and $C$ and have the
component expansions: 
\begin{equation}
\begin{array}{ccc}
B=\beta +i\theta _{+}b & \qquad \overline{B}=\overline{\beta }-i\theta _{-}%
\overline{b} &  \\ 
C=c+i\theta _{+}\gamma & \qquad \overline{C}=\overline{c}-i\theta _{-}%
\overline{\gamma } & 
\end{array}
\end{equation}
The supercurrents are: 
\begin{equation}
\begin{array}{cc}
j_{+}^{(\mathrm{ghost)}} & =-c\partial _{+}\beta +\frac 12\gamma b-\frac
32\partial _{+}c\beta \\ 
j_{-}^{(\mathrm{ghost)}} & =-\overline{c}\partial _{-}\overline{\beta }%
+\frac 12\overline{\gamma }\overline{b}-\frac 32\partial _{-}\overline{c}%
\overline{\beta }
\end{array}
\end{equation}
The fields $b,c,\beta ,\gamma $ satisfy the quantization conditions: 
\begin{equation}
\begin{array}{cc}
\left\{ \,b(\sigma ,\tau ),c({\sigma }^{\prime },\tau )\right\} \,=2\pi
\delta (\sigma -{\sigma }^{\prime }) &  \\ 
\left[ \beta (\sigma ,\tau ),\gamma ({\sigma }^{\prime },\tau )\right] =2\pi
\delta (\sigma -{\sigma }^{\prime }) & 
\end{array}
\end{equation}
One can define the ghost Dirac-Ramond operators by 
\begin{equation}
Q_{\pm }^{(\mathrm{ghost)}}=\frac 12\int d\sigma j_{\pm }^{(\mathrm{ghost)}}
\end{equation}
which will satisfy 
\begin{equation}
( Q_{\pm }^{(\mathrm{ghost)}} )^2=H^{(\mathrm{ghost})}\pm P^{(\mathrm{ghost})}
\end{equation}
The Hamiltonian and momenta of the ghost system do not interact with the
rest, but simply add up, allowing for this part to be computed seperately, as
it is independent of the background fields.

From the above analysis it should be clear that the operators $Q$ and $%
\overline{Q}$ togother with the Hilbert space and algebra of functions over
the loop space, define the geometry. The only other operators that we need
are the K-cycle involutions $\Gamma $ and $\overline{\Gamma }$ which satisfy 
\begin{equation}
\{\Gamma ,Q\}=0=[\overline{\Gamma },Q]
\end{equation}
and similarly for $\overline{Q}$: 
\begin{equation}
\{\overline{\Gamma },\overline{Q}\}=0=[\Gamma ,\overline{Q}]
\end{equation}
In the case of the superstring, these are the fermion numbers defined for
left and right movers (or here associated with $Q$ and $\overline{Q}$). From
reparametrization invariance one must insure that physical states satisfy $%
P=0$. The spectral action must be a function of of the operators $Q$ and $%
\overline{Q}$ as well as $\Gamma $ and $\overline{\Gamma }$. Therefore, and
on general grounds, a good candidate for the spectral action that will describe the
dynamics of the background fields is given by \cite{ACAC}
\begin{equation}
\mathrm{Tr\,\,}f\left( Q,\overline{Q},\Gamma ,\overline{\Gamma }\right)
\end{equation}
where the trace is taken over all states in the Hilbert space. Finding the
correct form of the function $f$ is a difficult task. However, even without
knowing the form of the function, the dependence on the background fields is
given in terms of geometric invariants. To every order in a heat-kernel type
expansion (through a Fourier or Melin transform) invariants would enter
multiplied by an overall numerical factor (Fourier components of the
function). This was the case when the spectral principle was applied to the
noncommutative space defining the standard model. There, the whole action
was determined, up to terms not higher than second order in curvature, in terms of
the first three geometric invariants in the heat kernel expansion. The
coefficients of these terms were chosen to fit known gauge and Higgs
couplings implying some relations among them.

Fortunately, in the situation considered here, the theory is known in two
limits. First when the background metric is flat, and the other is for the
zero modes where  one gets the low-energy field theory limit. First
we consider the case when the background geometry is flat.
 For flat
backgrounds, the Hamiltonian (\ref{hamilton}) simplifies to 
\beq
H=-\frac{1}{2T}\int d\sigma [ \frac{\delta}{\delta X^{\mu}}
\frac{\delta}{\delta X_{\mu}}-T^2 \frac{dX^{\mu}}{d\sigma}
\frac{dX_{\mu}}{d\sigma} -2iT(\chi_a\frac{d\chi^a}{d\sigma}+
\overline{\chi}^a \frac{d\overline{\chi}_a}{d\sigma})]
\eeq
This Hamiltonian could be expressed in terms of creation and
annihilation operators. 
 The path
integral expression of  the one-loop amplitude,is related to  
 the partition function \cite{Luis83}, in the case when the two-dimensional
 surface is a torus. The result is modular invariant, and therefore
consistent (free of anomalies) if the dimension of the target space
is ten. We also have to set $T=\frac{1}{4\pi l_s^2}$ where 
$l_s$ is the string length scale. Also to project the non-physical
states out (or equivalently, require modular invariance when the
two-dimensional surface has genus greater than or equal to two)
one must have the partition function \cite{SW}
\begin{equation}
I= \int \frac{d\tau d\overline{\tau}}{\tau_2^2} 
 \mathrm{Tr}\left| \sum_{NS\oplus R}\left( e^{2\pi
i(\tau Q^2}(-1)^\epsilon (1-\Gamma )
\right)\right|^2
\end{equation}
where $\epsilon =0$ for the NS sector, and $\epsilon =1$ for the Ramond
sector over the states in the trace. This action, has space-time
supersymmetry as can be verified by counting the number of fermionic and
bosonic states (massive as well as massless) and showing they are the same.
The parameter $\tau =\tau _1+i\tau _2$ is the modular parameter (although $%
\tau $ was used up to now as the two-dimensional time).
The total partition
function, including the ghosts is 
\begin{equation}
\int \frac{d\tau d\overline{\tau }}{\tau _2^2}
\int \frac{d^{8}p}{(2\pi )^8}e^{-2\pi p^2 \tau_2}
\left| \frac 1{2\eta (\tau
)^4}\left( \theta _3^4(0|\tau )-\theta _4^4(0|\tau )^4-\theta _2^4(0|\tau
)^4 -\theta _1^4(0|\tau )\right)  \right|^2
\label{thetas}
\end{equation}
and as expected, because of supersymmetry, the partition function vanishes.
The ghost contributions cancel the contributions of two bosonic and two
fermionic coordinates. Since the superghost part is independent of the
background, these contributions would be the same even in a curved
background. The difficulty is to compute the spectral action in an arbitrary
background including the dilaton, and the space-time supersymmetric vacuum
so that a space-time gravitino background, as well as a two and three forms
 would be included. This is not an easy problem to solve since this will
make space-time supersymmetry explicit without invoking the Green-Schwarz
superstring and $\kappa $ symmetry \cite{GSW}. For here we shall limit our
considerations to the background we started with (which is not the most
general, and can be perturbed to more general backgrounds by transformations
which are automorphisms of the algebra $\Omega (M)$).
To compute the spectral action in an arbitrary background is a very
complicated. We shall only determine the lowest order terms in a 
perturbative expansion. One starts by 
splitting the dependence of the fields in the  partition function in terms of 
zero modes and oscillators.
 In the NS-sector
there are no fermionic zero modes and the coordinates $X^{\mu}(\sigma )$ have
 a constant part $X_0^{\mu}$ . The Hamiltonian of the zero modes is
\beq
H_{\rm NS}^ 0=- [ \nabla_0^a \nabla_{0a} +\omega_{0b}^{\ \
  ab}\nabla_{0a}]
\eeq
 In the R-sector, there are fermionic zero modes
 $\chi_0^a $ and the zero modes Hamiltonian is
\begin{eqnarray*}
H_{\mathrm{R}}^0 &=&[-\nabla _0^a\nabla _{0a}+\omega _{0b}^{\ \  ab}\nabla
_{0a}+4\chi _0^a\bar \chi _0^b\chi _0^c\bar \chi _0^dR_{abcd}^0 \\
&&+\frac 23(\chi _0^a\bar \chi _0^b\bar \chi _0^c\bar \chi _0^d+\bar \chi
_0^a\chi _0^b\chi _0^c\chi _0^d+\bar \chi _0^a\chi _0^b\chi _0^c\bar \chi
_0^d)\nabla _{0a}H_{0bcd} \\
&&+2(\chi _0^b\chi _0^c+\bar \chi _0^b\bar \chi _0^c)H_{0bca}\nabla _{0a} \\
&&+\frac 13(\chi _0^a\chi _0^b\bar \chi _0^c\bar \chi _0^d+\chi _0^a\bar
\chi _0^b\bar \chi _0^c\chi _0^d+\chi _0^a\bar \chi _0^b\bar \chi _0^c\bar
\chi _0^d+\bar \chi _0^a\bar \chi _0^b\chi _0^c\chi _0^d)H_{0ab}^{\ \ e}
H_{0ecd}]\qquad 
\end{eqnarray*}
With these operators it is possible to use the heat kernel expansion to
evaluate the trace of the exponential
in the form \cite{Gilkey}
\[
\mathrm{Tr}(e^{-\tau_2 {\cal P}})=\sum_{n=0}^{\infty}a_n({\cal P})
\,\tau_2 ^{\frac{n-D}2}
\]
where $a_n({\cal P})$ are the Seeley-de Wit coefficients corresponding to the
operator ${\cal P}$ and $D=10$ is the dimension of the target manifold.
Using the results of heat kernel expansion for a general
second order operator, one finds the following results \cite{Gilkey}:
\[
\mathrm{Tr}(e^{-\tau_2 H_{\mathrm NS}^0})=\frac{a_0(H_{\mathrm
    NS}^0)}{\tau_2^5} +\frac{a_2(H_{\mathrm
    NS}^0)}{\tau_2^4}+\cdots 
\]
where
\beq
a_0(H_{\mathrm NS}^0)=\frac{1}{({2\pi})^5}\int d^{10}X_0 \sqrt{G[X_0]}
\eeq
and the center of mass coordinates $X_0^{\mu}$ become coordinates
on the manifold. The next term in the expansion is
\beq
a_2(H_{\mathrm NS}^0)=\frac{1}{({2\pi})^5}\int d^{10}X_0 \sqrt{G[X_0]}
( \frac{1}{6}R[X_0] )
\eeq
Similarly, for the R-sector, we have $a_0(H_{\mathrm R}^0)=
a_0(H_{\mathrm NS}^0)$ while for the next term $a_2$ we have

\beq
a_2(H_{\mathrm R}^0)=\frac{1}{({2\pi})^5}\int d^{10}X_0 \sqrt{G}(
-\frac{1}{12}R[X_0] -\frac{1}{24}H_{0\mu\nu\rho}H_{0}^{\mu\nu\rho}  )
\eeq
Higher orders in the expansion would involve higher curvature terms,
and will receive contributions from the oscillator parts. This can 
be done in a perturbative expansion using normal coordinates. To
lowest orders, and for the $a_0$ terms, this is given by an expansion
of the terms appearing in (\ref{thetas}) . This implies that the coefficient
of the $a_0$ term vanishes which is the cosmological constant.
For the $a_2 $ terms we have to expand , to lowest order in 
 $\tau $,  $(\theta_3^4 -\theta_4^4 )$
multiplying the NS-sector and $-(\theta_2^4 +\theta_1^4 )$ multiplying
the R-sector. The net contributions to lowest order is proportional to
\beq
\int d^{10}X_0 \sqrt{G[X_0]} ( \frac{1}{4}R[X_0] +\frac{1}{24}
H_{0\mu\nu\rho}H_{0}^{\mu\nu\rho} )
\eeq
Comparing this with the superstring effective action at low energies \cite{acdten, west}
we find that they are identical to this order.
 This is extremely encouraging, as we had no free
parameters to adjust. The challenging problem that remains is 
 to find a closed expression for
the spectral action as a function of the background geometry in 
analogy with the calculation of the elliptic genus in \cite{W87} where
modular invariance plays an important role.
 Of course the effective superstring action has
more terms depending on the dilaton, three-form, vector
and  gravitinos. It is possible to include the dilaton
by
adding to the non-linear sigma model the Weyl breaking term
\beq
\int d^2\xi d\theta_+d\theta_- \ {\rm sdet}E\  \phi[\Phi ]\ {\cal R}^{\pm}
\eeq
where $\phi $ is the background dilaton, and ${\cal R{\pm}}$ is the
super-curvature in two-dimensions and ${\rm sdet} E$ is the super-determinant
of the super-zweibein. As mentioned earlier, it is more difficult to
include in a curved background 
in a covariant way,   the spinors on the target
manifold.  The only known way is the Green-Schwarz formulation
which is studied in a light-cone gauge and is not explicitely
world-sheet
supesymmetric ( for an effective action derivation see e.g. \cite{Lerche}).
In a noncommutative formulation  it is 
quite important to have 
explicit world sheet supersymmetry as this is necessary to derive the
supercharges whose integrals give the Dirac-Ramond operators. These
points and other details are now under study.

In conclusion, we have shown that superstring non-linear sigma models
provide natural examples of noncommutative geometry spaces as 
developed by Connes. The tools of noncommutative geometry are
available
to study these spaces geometrically. Recent ideas proposed in
noncommutative geometry for  writing
a spectral action describing the dynamics of geometric fields (metric,
gauge fields, forms, \ldots ) are used and shown to give correct
answers in some known limits. It remains to find, in analogy with the case 
of the elliptic genus, closed form for this action in terms of
geometric
invariant. Details of the  results presented here  will 
appear somewhere else.

\end{document}